\documentclass[aps,prd,preprint,superscriptaddress,amsmath,amssymb,showpacs]{revtex4-1}
\usepackage{dcolumn}
\usepackage{graphicx}
\usepackage{float}
\usepackage{physics}
\usepackage[colorlinks=true,allcolors=blue]{hyperref}
\begin{document}
	\title{Spectral Signatures of Heavy Quarkonia in a Rotating and Anisotropic Quark-Gluon Plasma: A Holographic Study}
	\author{Xiang-Wei Shi}
	\affiliation{College of Mathematics and Physics, China Three Gorges University, Yichang 443002, China}
	
	\author{Sheng-Qin Feng}
	\email{Corresponding author: fengsq@ctgu.edu.cn}
	\affiliation{College of Mathematics and Physics, China Three Gorges University, Yichang 443002, China}
	\affiliation{Center for Astronomy and Space Sciences and Institute of Modern Physics, China Three Gorges University, Yichang 443002, China}
	\affiliation{Key Laboratory of Quark and Lepton Physics (MOE) and Institute of Particle Physics,\\
		Central China Normal University, Wuhan 430079, China}
	
	\date{\today}
	\begin{abstract}
		We investigate the in-medium spectral functions and effective masses of heavy quarkonia charmonium ($J/\Psi$) and bottomonium ($\Upsilon(1S)$) in a quark-gluon plasma (QGP) possessing both global rotation and spatial anisotropy. Using a gauge/gravity holographic model incorporating finite temperature, chemical potential, and warp factor, we compute the spectral signatures non-perturbatively. Our results show that both rotation and anisotropy enhance quarkonium dissociation, manifesting as peak suppression and width broadening in the spectral functions. Crucially, their effects are directional: anisotropy primarily dissociates longitudinally polarized states, while rotation more strongly disrupts transversely polarized ones. A competitive interplay exists: for small anisotropy, rotational effects dominate at high angular velocity, whereas for large anisotropy, anisotropy governs the dissociation regardless of rotation strength. Furthermore, rotation induces a non-monotonic temperature dependence in the transverse effective mass of $J/\Psi$, while strong anisotropy causes similar non-monotonicity in the longitudinal effective mass of $J/\Psi$. These findings reveal how the distinct symmetry breaking patterns induced by QGP rotation and anisotropy reshape the heavy quarkonium spectrum, providing new insights into polarization-dependent suppression in non-central heavy-ion collisions.
   \end{abstract}
   \maketitle
\section{Introduction}\label{sec:Intro}
Ultra-relativistic heavy-ion collisions occur in particle accelerators creating a transient state of matter known as quark-gluon plasma (QGP)~\cite{PHENIX:2004vcz,STAR:2005gfr,BRAHMS:2004adc}, which exhibits nearly perfect fluid behavior. Due to its extremely short lifetime ($\sim5-\/10  fm/c$), the QGP cannot be observed directly. Heavy mesons, such as those composed of charm ($\textit{c}$\,) and bottom (b) quarks, survive longer in the hot medium than light mesons made of up (u), down (d), and strange (s) quarks, and undergo a temperature-dependent and density-dependent partial dissociation process. Thus, studying heavy quarkonia provides a valuable probe of the internal properties of the QGP.

A detailed description of the quantum chromodynamics (QCD) phase diagram, encompassing parameters such as temperature, chemical potential, magnetic field strength, among others, remains a formidable yet critically important challenge in high-energy physics. In the region of strongly coupled quantum chromodynamics (QCD), the traditional perturbation method is no longer applicable, and lattice QCD also faces well-known difficulties in the presence of a finite chemical potential. Consequently, investigating the strongly coupled quark-gluon plasma produced in heavy-ion collisions at facilities like RHIC and the LHC requires non-perturbative approaches. In recent years, gauge/gravity duality (holography) has emerged as a powerful non-perturbative tool for studying QCD~\cite{Deng:2021kyd,Chen:2017lsf,Chen:2020ath,Zhu:2019ujc} and has been widely applied to investigate heavy-flavor dynamics~\cite{Zhao:2021ogc,Fujita:2009ca,Kim:2007rt,Braga:2017bml,Gubser:2007zr,Finazzo:2014rca,Hashimoto:2014fha,Lee:2013oya,Fadafan:2013coa,BitaghsirFadafan:2015zjc,BitaghsirFadafan:2015yng,Braga:2020myi,Iatrakis:2015sua,Braga:2016oem,Bellantuono:2017msk,Feng:2019boe,Zollner:2020nnt}.

In non-central ultra-relativistic heavy-ion collisions, not only are intense magnetic fields generated~\cite{Guo:2019joy,She:2017icp,Zhong:2014cda,Bzdak:2011yy,Deng:2012pc}, but also considerable angular momentum$-\/$reaching magnitudes of order ${10^4} - {10^5}$$\hbar$ with local angular velocities in the range 0.01$-\/$0.1 GeV~\cite{Liang:2004ph,Huang:2011ru,Becattini:2007sr,Jiang:2016wvv,Pang:2016igs,Csernai:2013bqa,Deng:2016gyh}. Numerous holographic studies~\cite{Zhao:2023pne,Wang:2024rim,Wang:2024szr} and effective field theory treatments~\cite{Sun:2021hxo,Qiu:2023kwv,Qiu:2023ezo,Bao:2024glw,Zhu:2022irg}~have incorporated rotational effects. For example, Ref.~\cite{Zhao:2023pne} examined the combined influence of temperature, chemical potential, rotation, and magnetic field on heavy vector meson dissociation, revealing a competition among these factors. Ref.~\cite{Wang:2024rim} studied the role of rotational radius and found that larger radii enhance dissociation, with a more pronounced effect perpendicular to the direction of rotational angular velocity.

Introducing anisotropy into holographic models is essential~\cite{Janik:2008tc,Mateos:2011ix,Mateos:2011tv,Rebhan:2012bw,Arefeva:2014vjl,Gursoy:2018ydr}, because the QGP exhibits significant  anisotropy shortly after a high-energy heavy-ion collision, with an estimated isotropization time of about 1$-\/$5 ~$fm/c$ ~\cite{Strickland:2013uga}. One motivation for anisotropic models~\cite{Arefeva:2014vjl} is to reproduce the energy dependence of particle multiplicities observed in experiments~\cite{ALICE:2015juo}.

Many previous holographic studies of heavy vector mesons have assumed an isotropic, homogeneous QGP medium. However, experimental evidence from heavy-ion collisions indicates that the early-stage QGP possesses a strong local anisotropy, with the system expanding predominantly along the collision axis~\cite{Giataganas:2013lga}. Reference~\cite{Chang:2024ksq}~investigated the effect of an anisotropic QGP on heavy vector meson dissociation and found that increased anisotropy promotes dissociation, with a stronger effect for mesons polarized parallel to the anisotropy direction. Motivated by these results, the present work explores how a QGP possessing both global rotation and spatial anisotropy influences the spectral functions of heavy vector mesons.

This paper is organized as follows. Section II introduces the anisotropic five-dimensional black-brane solution. Section III details the calculation of spectral functions after introducing rotation. Numerical results and discussion are presented in Section IV. Finally, a summary and conclusions are given in Section V.
\section{ANISOTROPIC HOLOGRAPHIC MODEL }\label{sec:2}
In this section, we review the anisotropic holographic model proposed in~\cite{Arefeva:2018hyo} with an arbitrary dynamical exponent $\nu$. Based on the five-dimensional Einstein-dilaton-two-Maxwell system, the action of the system can be specified as
\begin{equation}\label{eq:1}
	\begin{aligned}
	S = \int {\frac{{{d^5}x}}{{16\pi {G_5}}}\sqrt { - \det ({g_{\mu \upsilon }})} \left[ {R - \frac{{{f_1}(\phi )}}{4}{F^2}_{(1)} - \frac{{{f_2}(\phi )}}{4}{F^2}_{(2)} - \frac{1}{2}{\partial _\mu }\phi {\partial ^\mu }\phi  - V(\phi )} \right]} ,
    \end{aligned}
\end{equation}
where ${F^2}_{(1)}$ and ${F^2}_{(2)}$ are the squares of the Maxwell field, with $F_{\mu \upsilon }^{(1)} = {\partial _\mu }{A_\upsilon } - {\partial _\upsilon }{A_\mu }$~and $F_{\mu \upsilon }^{(2)} = q{\kern 1pt} d{y^1} \wedge q{\kern 1pt} d{y^2}$,~respectively. The first Maxwell field induces finite chemical potential, the second Maxwell field provides the spatial anisotropy. ${f_1}(\phi )$ and ${f_2}(\phi )$ are the gauge kinetic functions related to the corresponding Maxwell fields. $V(\phi )$ is the potential of the scalar field $\phi $. We use label~ ${\tilde x^\mu } = (\tilde t,\tilde z,{\tilde x_1},{\tilde x_2},{\tilde x_3})$ to denote the stationary frame. The form of metric ansatz is adopted as
\begin{equation}\label{eq:2}
	\begin{aligned}
		d{\tilde{s}^2} = \frac{\tilde{L}^2b(\tilde{z})}{{{\tilde{z}^2}}}\left[ { - g(\tilde{z})d{\tilde{t}^2} + d{\tilde{x}_1}^2 + {\tilde{z}^{2 - \frac{2}{\nu }}}(d{\tilde{x}_2}^2 + d{\tilde{x}_3}^2) + \frac{{d{\tilde{z}^2}}}{{g(\tilde{z})}}} \right] ,
	\end{aligned}
\end{equation}
\begin{equation}\label{eq:3}
	\begin{aligned}
		\phi  = \phi (\tilde{z}),\;A_\mu ^{(1)} = {A_t}(\tilde{z})\delta _\mu ^0  ~,
	\end{aligned}
\end{equation}
where $g(\tilde{z})$ is the blackening function, $b(\tilde{z}) = {e^{\frac{1}{2}c{\tilde{z}^2}}}$ is the AdS deformation factor, with the warp factor coefficient $\textit{c}$ signifying the deviation from conformality, and the AdS radius $\tilde{L} = 1 ~\textrm{GeV}^{ - 1}$ is set for convenience. By solving the motion equations derived from the action mentioned above, one can obtain the blackening function as
\begin{equation}\label{eq:4}
	\begin{aligned}
		g(\tilde z) = 1 - \frac{{{{\tilde z}^{2 + \frac{2}{\nu }}}}}{{{{\tilde z}_h}^{2 + \frac{2}{\nu }}}}\frac{{\mathfrak{G}\left( {\frac{3}{4}c{{\tilde z}^2}} \right)}}{{\mathfrak{G}\left( {\frac{3}{4}c{{\tilde z}_h}^2} \right)}} - \frac{{{{\tilde \mu }^2}c{{\tilde z}^{2 + \frac{2}{\nu }}}{e^{\frac{{c{{\tilde z}_h}^2}}{2}}}}}{{4{{\left( {1 - {e^{\frac{{c{{\tilde z}_h}^2}}{4}}}} \right)}^2}}}\mathfrak{G}\left( {c{{\tilde z}^2}} \right) + \frac{{{{\tilde \mu }^2}c{{\tilde z}^{2 + \frac{2}{\nu }}}{e^{\frac{{c{{\tilde z}_h}^2}}{2}}}}}{{4{{\left( {1 - {e^{\frac{{c{{\tilde z}_h}^2}}{4}}}} \right)}^2}}}\frac{{\mathfrak{G}\left( {\frac{3}{4}c{{\tilde z}^2}} \right)}}{{\mathfrak{G}\left( {\frac{3}{4}c{{\tilde z}_h}^2} \right)}}\mathfrak{G}\left( {c{{\tilde z}_h}^2} \right) ~,
	\end{aligned}
\end{equation}
where
\begin{equation}\label{eq:5}
	\begin{aligned}
		\mathfrak{G}\left( x \right) = \sum\limits_{n\, = \,0}^\infty  {\frac{{{{\left( { - 1} \right)}^n}{x^n}}}{{n!\left( {1 + n + \frac{1}{\nu }} \right)}}} ~,
	\end{aligned}
\end{equation}
where the anisotropic parameter $\nu $ is restricted within a range of  ($1 < \nu  \le 4.5$) to satisfy the dependence of the particle multiplicity on energy of ALICE experiment of LHC energy region~\cite{ALICE:2015juo}.  All studies~\cite{Alvarez-Gaume:2008qeo,Lin:2009pn,Albacete:2009ji,Arefeva:2009oun,Arefeva:2009pxq,Kovchegov:2009du,Kovchegov:2010zg,Kiritsis:2011yn,Arefeva:2013fia} using isotropic models failed in reproducing the dependence of the particle multiplicity on energy.
Furthermore, the temperature is obtained by the following formula as
\begin{equation}\label{eq:6}
	\begin{aligned}
		\tilde {T} \left( {\tilde{z}_h,\tilde{\mu} ,c,\nu } \right) =\frac{g'(\tilde{z}_h)}{4\pi} = \frac{{{e^{ - \frac{3}{4}c{\tilde{z}_h}^2}}}}{{2\pi {\tilde{z}_h}}}\left| {\frac{1}{{\mathfrak{G}\left( {\frac{3}{4}c{\tilde{z}_h}^2} \right)}} + \frac{{{\tilde{\mu} ^2}c{\tilde{z}_h}^{2 + \frac{2}{\nu }}{e^{\frac{{c{\tilde{z}_h}}}{4}}}}}{{4{{\left( {1 - {e^{\frac{{c{\tilde{z}_h}^2}}{4}}}} \right)}^2}}}\left( {1 - {e^{\frac{{c{\tilde{z}_h}^2}}{4}}}\frac{{\mathfrak{G}\left( {c{\tilde{z}_h}^2} \right)}}{{\mathfrak{G}\left( {\frac{3}{4}c{\tilde{z}_h}^2} \right)}}} \right)} \right|  ,
	\end{aligned}
\end{equation}
where the temperature function $\tilde{T}({\tilde{z}_h})$ is multivalued when the chemical potential is below its critical value as $\tilde{\mu}  < {\tilde{\mu} _{cr}}$, and becomes single-valued when $\tilde{\mu}  \geq {\tilde{\mu} _{cr}}$  ( For a more detailed discussion of this model, please refer to~\cite{Arefeva:2018hyo} ). Based on ~\cite{Zhou:2021sdy,Braga:2018zlu,Nadi:2019bqu}, the holographic model is expanded to the situation of rotation with a planar horizon. The general metric in the rest frame is given by
\begin{equation}\label{eq:7}
	\begin{aligned}
		d{\tilde s^2} =  - {\tilde g_{tt}}d{\tilde t^2} + {\tilde g_{{x_1}{x_1}}}d{\tilde x_1}^2 + {\tilde g_{{x_{2,3}}{x_{2,3}}}}d\tilde x_{2,3}^2 + {\tilde g_{zz}}d{\tilde z^2} .
	\end{aligned}
\end{equation}

In order to analyze the influence of the rotating QGP,  the above metric should be manifested in a cylindrical symmetric form
\begin{equation}\label{eq:8}
	\begin{aligned}
		d{\tilde s^2} =  - {\tilde g_{tt}}d{\tilde t^2} + {\tilde g_{{x_1}{x_1}}}{l^2}d{\tilde \theta ^2} + {\tilde g_{{x_{2,3}}{x_{2,3}}}}d\tilde x_{2,3}^2 + {\tilde g_{zz}}d{\tilde z^2}.
	\end{aligned}
\end{equation}

It should be noted that we use label ~${x^\mu } = (t,z,{x_1},{x_2},{x_3})$ to denote the rotating frame. The angular momentum can be turned on in the angular coordinate $\tilde \theta $ through the standard Lorentz transformation
\begin{equation}\label{eq:9}
	\begin{aligned}
		\tilde \theta  \to \gamma \,\left( {\theta  + \omega \,\,t} \right),\; \tilde t \to \gamma \left( {t + \omega \,{l^2}\,\theta } \right),
	\end{aligned}
\end{equation}
where $\gamma  = \frac{1}{{\sqrt {1 - {\omega ^2}{l^2}} }}$ is the usual Lorentz factor, $\omega $ is the angular velocity of rotation, $l$ is the radius of the rotating axis (The discussions in the following text all assume $l = 1 ~\textrm{GeV}{^{ - 1}}$). The above metric will be transformed into the following form
\begin{equation}\label{eq:10}
	\begin{aligned}
	 d{s^2} = \,&{\gamma ^2}({\tilde g_{{x_1}{x_1}}}{\omega ^2}{l^2} - {\tilde g_{tt}})d{t^2} + 2{\gamma ^2}\omega {l^2}({\tilde g_{{x_1}{x_1}}} - {\tilde g_{tt}})dtd\theta \\& + {\gamma ^2}({\tilde g_{{x_1}{x_1}}} - {\omega ^2}{l^2}{\tilde g_{tt}}){l^2}d{\theta ^2}  + {\tilde g_{zz}}d{z^2} + {\tilde g_{{x_{2,3}}}}dx_{2,3}^2 .	
	\end{aligned}
\end{equation}

The Hawking temperature and chemical potential of the rotating black hole can be given as
\begin{equation}\label{eq:11}
	\begin{aligned}
	T = \tilde T({\tilde z_h},\tilde \mu ,c,\nu )\sqrt {1 - {\omega ^2}{l^2}},\; ~~\mu  = \tilde \mu \sqrt {1 - {\omega ^2}{l^2}} .
	\end{aligned}
\end{equation}
\section{THE SPECTRAL FUNCTIONS }\label{sec:3}
In this section, we will calculate the spectral function of the heavy vector mesons after the introduction of the rotational effect. In order to go smoothly later, we redefine the metric from Eq.\eqref{eq:10} as
\begin{equation}\label{eq:12}
	\begin{aligned}
	d{s^2} =  - {g_{tt}}d{t^2} + {g_{t{x_1}}}dtd{x_1} + {g_{{x_1}t}}d{x_1}dt + {g_{{x_1}{x_1}}}d{x_1}^2 + {g_{{x_2}{x_2}}}d{x_2}^2 + {g_{{x_3}{x_3}}}d{x_3}^2 + {g_{zz}}d{z^2} .	
	\end{aligned}
\end{equation}

The standard Maxwell action is
\begin{equation}\label{eq:13}
	\begin{aligned}
	S =  - \int {{d^4}xdz\frac{Q}{4}{F_{mn}}{F^{mn}}},	
	\end{aligned}
\end{equation}
where $Q = \frac{{\sqrt { - g} }}{{h(\phi ){g_5}^2}},$ $h\left( \phi  \right) = {e^{\phi (z)}},$ ${F_{mn}} = {\partial _m}{V_n} - {\partial _n}{V_m},$  and the vector field ${V_m} = \left( {{V_\mu },{V_z}} \right)$ $\left( {\mu  = 0,1,2,3} \right)$ is introduced to represent the heavy vector mesons, which is dual to the gauge theory current ${J^\mu } = \mathop \Psi \limits^\_ {\gamma ^\mu }\Psi $.  $\phi \left( z \right)$ is the dilaton background
\begin{equation}\label{eq:14}
	\begin{aligned}
	\phi \left( z \right) = {\kappa ^2}{z^2} + Mz + \tanh \left( {\frac{1}{{Mz}} - \frac{\kappa }{{\sqrt \Gamma  }}} \right),	
	\end{aligned}
\end{equation}
where $\kappa ,\Gamma \,\;{\rm{and}}\;M$ are for the quark mass, the string tension, and a large mass related to the heavy quarkonium non-hadronic decay, respectively. By fitting the mass spectrum, one can obtain the values of the three energy parameters for charmonium and bottomonium ~\cite{Mamani:2022qnf}, respectively as
\begin{equation}\label{eq:15}
	\begin{aligned}
	{\kappa _c} = 1.2\;{\text{GeV}},\;\sqrt {{\Gamma _c}}  = 0.55\;{\text{GeV}},\;{M_c} = 2.2\;{\text{GeV}},
	\end{aligned}
\end{equation}
\begin{equation}\label{eq:16}
	\begin{aligned}
	{\kappa _b} = 2.45\;{\text{GeV}},\;\sqrt {{\Gamma _b}}  = 1.55\;\,{\text{GeV}},\;{M_b} = 6.2\,{\text{GeV}} .
	\end{aligned}
\end{equation}

The spectral functions of the heavy vector mesons can be calculated using the membrane paradigm ~\cite{Iqbal:2008by}. The motion equations can be obtained from Eq.\eqref{eq:13} as
\begin{equation}\label{eq:17}
	\begin{aligned}
	{\partial _m}(Q{F^{mn}}) = {\partial _z}(Q{F^{zn}}) + {\partial _\mu }(Q{F^{\mu n}})~,	
	\end{aligned}
\end{equation}
where ${F^{mn}} = {g^{m\alpha }}{g^{n\beta }}{F_{\alpha \beta }},$ $n = \left( {0,1,2,3,4} \right),$ $\mu  = \left( {0,1,2,3} \right).$ For the $z$-foliation, the conjugate momentum of the gauge field is given by
\begin{equation}\label{eq:18}
	\begin{aligned}
	{j^\mu } =  - Q{F^{z\mu }} .
	\end{aligned}
\end{equation}

The propagation direction of the plane wave solution of the vector field is assumed to be along ${x_1}$, which is the same as the anisotropic direction. The motion equation can be decomposed into two parts: the longitudinal channel and the transverse channel. The first involves the fluctuations along the $(t,{x_1})$ direction, the second involves the fluctuations along the $({x_2},{x_3})$. From Eq.\eqref{eq:17}, one can obtain the dynamical equation of the longitudinal channel as
\begin{equation}\label{eq:19}
	\begin{aligned}
	 - {\partial _z}{j^t} - \frac{{\sqrt { - g} }}{{h\left( \phi  \right)}}({g^{{x_1}{x_1}}}{g^{tt}} + {g^{{x_1}t}}{g^{t{x_1}}}){\partial _{{x_1}}}{F_{{x_1}t}} = 0 ,	
	\end{aligned}
\end{equation}
\begin{equation}\label{eq:20}
	\begin{aligned}
	- {\partial _z}{j^{{x_1}}} + \frac{{\sqrt { - g} }}{{h\left( \phi  \right)}}({g^{tt}}{g^{{x_1}{x_1}}} + {g^{t{x_1}}}{g^{{x_1}t}}){\partial _t}{F_{{x_1}t}} = 0 ,	
	\end{aligned}
\end{equation}
\begin{equation}\label{eq:21}
	\begin{aligned}
	{\partial _{{x_1}}}{j^{{x_1}}} + {\partial _t}{j^t} = 0 .	
	\end{aligned}
\end{equation}

From Bianchi's identity, one can obtain:
\begin{equation}\label{eq:22}
	\begin{aligned}
	{\partial _z}{F_{{x_1}t}} - \frac{{{g_{zz}}h\left( \phi  \right)}}{{\sqrt { - g} }}{\partial _t}\left[ {{g_{{x_1}{x_1}}}{j^{{x_1}}} + {g_{{x_1}t}}{j^t}} \right] - \frac{{{g_{zz}}h\left( \phi  \right)}}{{\sqrt { - g} }}{\partial _{{x_1}}}\left[ {{g_{tt}}{j^t} - {g_{t{x_1}}}{j^{{x_1}}}} \right] = 0 .	
	\end{aligned}
\end{equation}

The conductivity of the longitudinal channel and its derivative with respect to $z$ can be expressed in the following form
\begin{equation}\label{eq:23}
	\begin{aligned}
	{\sigma _L}(\varepsilon ,z) = \frac{{{j^{{x_1}}}(\varepsilon ,z)}}{{{F_{{x_1}t}}(\varepsilon ,z)}} ,	
	\end{aligned}
\end{equation}
\begin{equation}\label{eq:24}
	\begin{aligned}
	{\partial _z}{\sigma _L}(\varepsilon ,z) = \frac{{{\partial _z}{j^{{x_1}}}}}{{{F_{{x_1}t}}}} - \frac{{{j^{{x_1}}}}}{{F_{{x_1}t}^2}}{\partial _z}{F_{{x_1}t}} ~.
	\end{aligned}
\end{equation}

According to the Kubo's formula, one can obtain that the conductivity at the boundary is related to the retarded Green's function as
\begin{equation}\label{eq:25}
	\begin{aligned}
	{\sigma _L}(\varepsilon ) = \frac{{ - G_R^L(\varepsilon )}}{{i\varepsilon }} .	
	\end{aligned}
\end{equation}

In order to obtain the specific form of the aforementioned flow equation, we assume ${A_\mu } = {A_n}(p,\,z)\,{e^{ - i\varepsilon t + ip{x_1}}},$ where ${A_n}(p,z)$ is the quasi-normal modes. So one can obtain ${\partial _t}{F_{{x_1}t}} =  - i\varepsilon {F_{{x_1}t}},$ ${\partial _t}{j^{{x_1}}} =  - i\varepsilon {j^{{x_1}}}.$ From Eqs.\eqref{eq:20} \eqref{eq:21} and \eqref{eq:22} and taking the momentum limit, Eq.\eqref{eq:24} can be obtained as
\begin{equation}\label{eq:26}
	\begin{aligned}
	{\partial _z}{\sigma _L}(\varepsilon ,z) = i\varepsilon {\Delta _L}\left( {\sigma _L}{{\left( {\varepsilon ,z} \right)}^2} - {\Sigma_{L} ^2} \right),	
	\end{aligned}
\end{equation}
where
\begin{equation}\label{eq:27}
	\begin{aligned}
	{\Delta _L} = \frac{{\left( {1 - {\omega ^2}{l^2}g(z)} \right)\sqrt {b(z)} {e^{\phi (z)}}}}{{g(z)(1 - {\omega ^2}{l^2}){z^{1 - \frac{2}{\nu }}}}},	
	\end{aligned}
\end{equation}
and
\begin{equation}\label{eq:28}
	\begin{aligned}
	\Sigma_{L}^{2} = \frac{{b(z){e^{ - 2\phi (z)}}{z^{2 - \frac{4}{\nu }}}(1 - {\omega ^2}{l^2})}}{{ {1 - {\omega ^2}{l^2}g(z)} }} .
	\end{aligned}
\end{equation}

The dynamical equation for the transverse channel can be derived as
\begin{equation}\label{eq:29}
	\begin{aligned}
	 \partial_{z}j^{x_{2}} &- \frac{\sqrt{-g}}{h(\phi)} \Bigl[
	 g^{tx_{1}}g^{x_{2}x_{2}}\partial_{t}F_{x_{1}x_{2}}
	- g^{tt}g^{x_{2}x_{2}}\partial_{t}F_{tx_{2}} \\
	& + g^{x_{1}t}g^{x_{2}x_{2}}\partial_{x_{1}}F_{tx_{2}}
	+ g^{x_{1}x_{1}}g^{x_{2}x_{2}}\partial_{x_{1}}F_{x_{1}x_{2}} \Bigr] = 0 ,	
	\end{aligned}
\end{equation}
\begin{equation}\label{eq:30}
	\begin{aligned}
	\frac{{{g_{{x_2}{x_2}}}{g_{zz}}h\left( \phi  \right)}}{{\sqrt { - g} }}{\partial _t}{j^{{x_2}}} + {\partial _z}{F_{t{x_2}}} = 0 ,	
	\end{aligned}
\end{equation}
\begin{equation}\label{eq:31}
	\begin{aligned}
	{\partial _{{x_1}}}{F_{t{x_2}}} + {\partial _t}{F_{{x_2}{x_1}}} = 0 .	
	\end{aligned}
\end{equation}

The transverse conductivity and its derivative are obtained as
\begin{equation}\label{eq:32}
	\begin{aligned}
	{\sigma _T}\left( {\varepsilon ,z} \right) = \frac{{{j^{{x_2}}}\left( {\varepsilon ,\vec p,z} \right)}}{{{F_{{x_2}t}}\left( {\varepsilon ,\vec p,z} \right)}} ,	
	\end{aligned}
\end{equation}
\begin{equation}\label{eq:33}
	\begin{aligned}
	{\partial _z}{\sigma _T}\left( {\varepsilon ,z} \right) = \frac{{{\partial _z}{j^{{x_2}}}}}{{{F_{{x_2}t}}}} - \frac{{{j^{{x_2}}}}}{{F_{{x_{2t}}}^2}}{\partial _z}{F_{{x_2}t}} .	
	\end{aligned}
\end{equation}

Similarly, we have ${\partial _t}{F_{{x_1}{x_2}}} =  - i\varepsilon {F_{{x_1}{x_2}}},{\partial _t}{F_{t{x_2}}} =  - i\varepsilon {F_{t{x_2}}},{\partial _t}{j^{{x_2}}} =  - i\varepsilon {j^{{x_2}}}$. Therefore, the transverse flow equation can be obtained as
\begin{equation}\label{eq:34}
	\begin{aligned}
	{\partial _z}{\sigma _T}(\varepsilon ,z) = i\varepsilon {\Delta _T}\left( {\sigma _T}{{\left( {\varepsilon ,z} \right)}^2} - {\Sigma_{T} ^2} \right) ,
	\end{aligned}
\end{equation}
where
\begin{equation}\label{eq:35}
	\begin{aligned}
	{\Delta _T} = \frac{{z{e^{\phi (z)}}}}{{g(z)\sqrt {b(z)} }} ,	
	\end{aligned}
\end{equation}
and
\begin{equation}\label{eq:36}
	\begin{aligned}
	\Sigma_{T}^{2} = \frac{{b(z){e^{ - 2\phi (z)}}(1 - {\omega ^2}{l^2}g(z))}}{{{z^2}(1 - {\omega ^2}{l^2})}} .	
	\end{aligned}
\end{equation}

The above equation can be calculated by using the initial conditions, which can be obtained by requiring regularity at the horizon ${\partial _z}{\sigma _{L(T)}}(\varepsilon ,z) = 0.$ Eqs.\eqref{eq:26} and \eqref{eq:34} have the same form, when the angular velocity is 0 and the anisotropic parameter is 1. The spectral function can be defined by the retarded Green's function as
$\rho \left( \varepsilon  \right) \equiv  - \operatorname{Im} {G_R}\left( \varepsilon  \right) = \varepsilon \operatorname{Re} \sigma \left( {\varepsilon ,0} \right)$.
\section{NUMERICAL RESULTS OF SPECTRAL FUNCTIONS IN A BACKGROUND OF ROTATION AND ANISOTROPY }\label{sec:4}
After introducing rotation and anisotropy, the spectral function of the heavy vector mesons can be calculated using the membrane paradigm~\cite{Iqbal:2008by}. In this section, we first investigate the influence of chemical potential and temperature on the spectral function of the heavy vector mesons under the anisotropic and rotating background. Next, we considered the combined effect of rotation and anisotropy on the spectral function. Finally, we investigated the effects of anisotropy and rotational effects on the effective mass of the heavy vector mesons.

\subsection{THE INFLUENCE OF CHEMICAL POTENTIAL AND TEMPERATURE ON THE SPECTRAL FUNCTIONS }\label{sec:A}
\begin{figure}[H]
	\centering
	\includegraphics[width=0.61\textwidth]{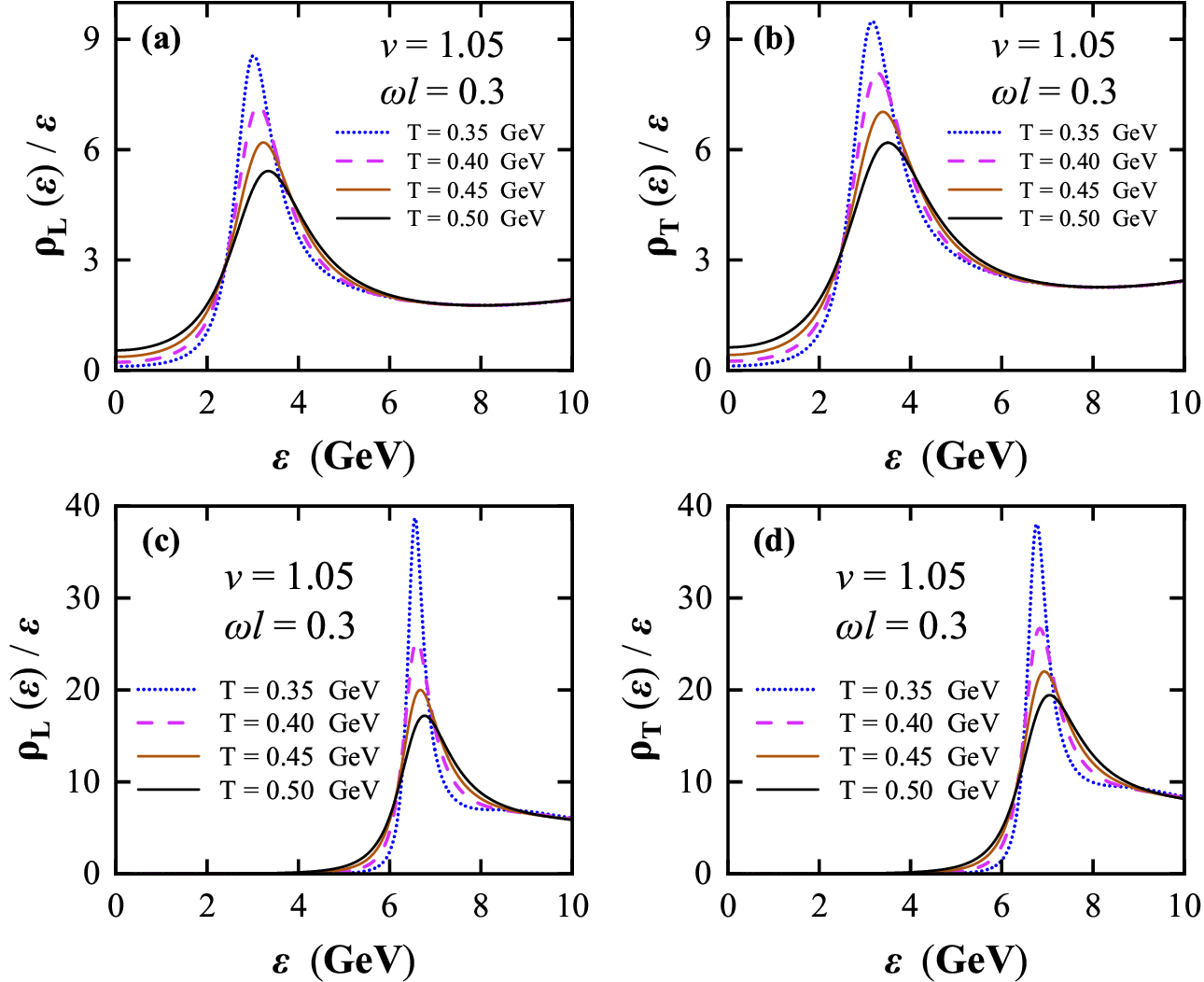}
    \caption{Spectral functions of $J/\Psi $ and $\Upsilon (1S)$ at different temperatures, with fixed anisotropy parameter $\nu  = 1.05$, $\mu  = 0.1\,{\rm{GeV}},c =  - 0.3\;{\rm{Ge}}{{\rm{V}}^2}$ and $\omega  = 0.3~{\rm{GeV}}$. Panels (a) and (b) correspond to $J/\Psi $, while (c) and (d) correspond to $\Upsilon (1S)$. The left panels (a, c) show the spectral function parallel to the anisotropic direction (i.e., the direction of the rotational angular velocity), and the right panels (b, d) show the spectral function perpendicular to it.}
	\label{fig:1}
\end{figure}
Fig.~\ref{fig:1} presents the temperature dependence of the spectral functions for the heavy vector mesons $J/\Psi $ and $\Upsilon (1S)$ in an anisotropic and rotating QGP background. The bell shape spectral functions represents the resonance state in the medium. The peak location reflects the in-medium effective mass of the state, the width corresponding to half the peak height is the resonant decay width. A narrower width indicates a longer-lived quasi-particle state. For both $J/\Psi $ and $\Upsilon (1S)$, increasing temperature leads to a systematic suppression of the spectral peak height and a broadening of the peak width. This behavior is observed in both longitudinal (parallel to anisotropy) and transverse (perpendicular to anisotropy) polarizations. The suppression and broadening are clear signatures of enhanced dissociation, indicating that higher thermal energy disrupts the heavy quarkonium bound states, promoting their melting into the deconfined plasma.

Although temperature affects both polarization directions, a careful comparison reveals that the longitudinal spectral peaks ( parallel to the anisotropy ) show marginally greater suppression and broadening than the transverse ones, especially for $J/\Psi $. This suggests that the anisotropic and rotating background cause a directional effect on the dissociation process ( For a comprehensive exposition of the research, please refer to the subsequent section ). The anisotropy and rotation in the medium breaks spatial symmetry, affecting meson states differently which depends on their different polarization directions. The bottomonium ($\Upsilon (1S)$) peaks are noticeably taller and narrower than those of charmonium ($J/\Psi $) at comparable temperatures. This reflects the larger bottom quark mass can lead to a stronger binding energy and greater resistance to thermal dissociation. The result is consistent with the established melting picture, where heavier quarkonia survive to higher temperatures.

The observed spectral changes mirror the expected behavior of quarkonium suppression in the hot, anisotropic and rotating QGP created in non-central relativistic heavy-ion collisions. The dissociation of heavy quarkonium exhibits the dependence of derection, which may stem from the combined effects of rotation and anisotropy. Fig.~\ref{fig:1} establishes the foundational thermal behavior of heavy vector mesons in an anisotropic and rotating backgroung. It confirms that temperature universally enhances dissociation while revealing subtle directional dependencies induced by anisotropy and rotation. The findings underscore the importance of including anisotropic and rotation effects in theoretical models to more precisely interpret quarkonium suppression data from experiments such as those at the LHC and RHIC.
\begin{figure}[H]
	\centering
	\includegraphics[width=0.61\textwidth]{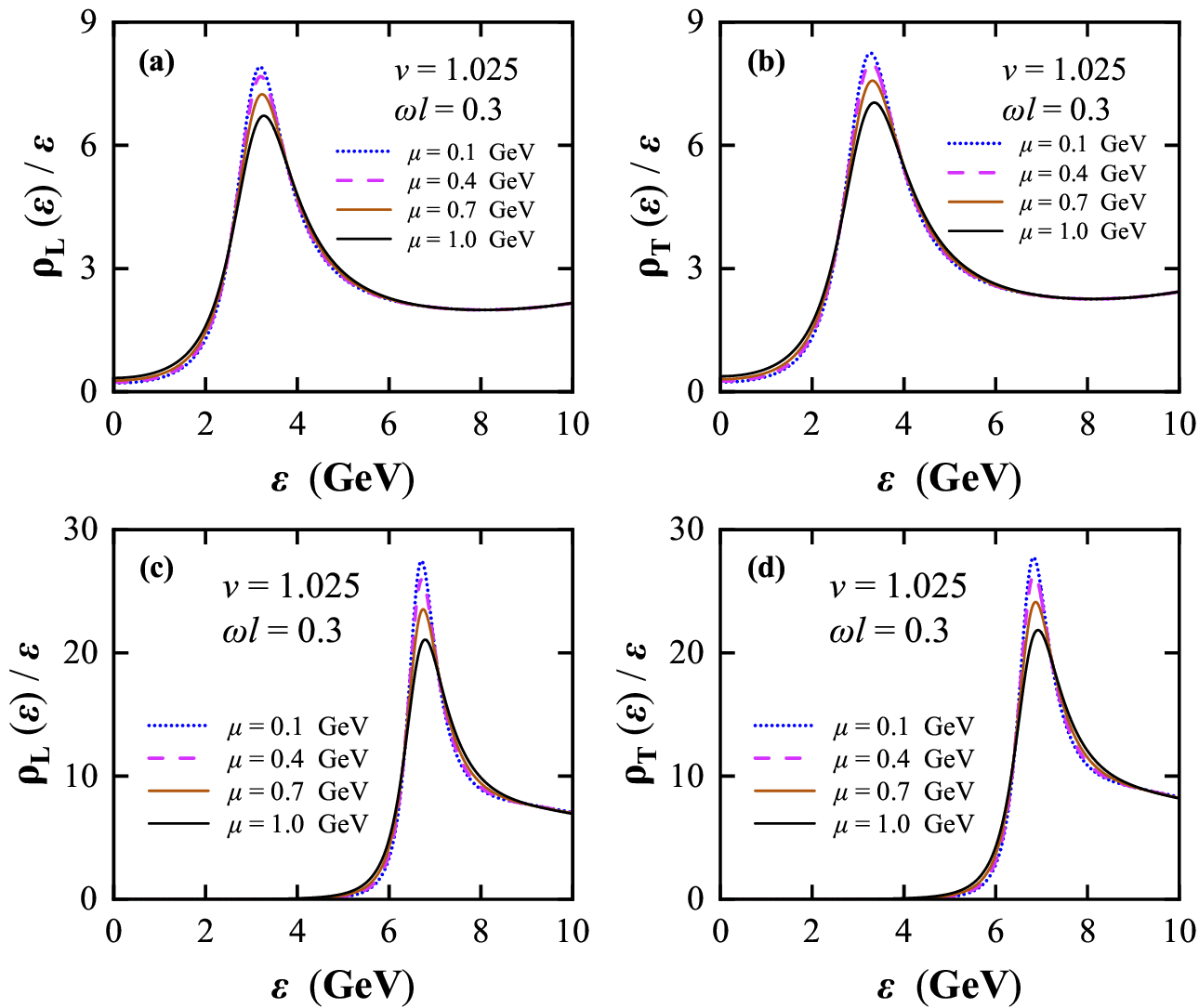}
    \caption{Spectral functions of $J/\Psi $ and $\Upsilon (1S)$ at different chemical potentials ($\mu $), with fixed temperature $T$ = 0.4 GeV and $c =  -0.3\;{\text{Ge}}{{\text{V}}^2}$. Panels (a) and (b) correspond to $J/\Psi $ , while (c) and (d) correspond to $\Upsilon (1S)$. The left panels (a, c) show the spectral function parallel to the anisotropic direction, and the right panels (b, d) show the spectral function perpendicular to it.}
	\label{fig:2}
\end{figure}
Fig.~\ref{fig:2} systematically examines the influence of baryon chemical potential ($\mu $) on the spectral functions of $J/\Psi $ and $\Upsilon (1S)$ within an anisotropic and rotating QGP, at a fixed temperature. Increasing the chemical potential leads to a clear suppression of the spectral peak height and a concomitant broadening of the peak width for both $J/\Psi $ and $\Upsilon (1S)$. This behavior is consistent across both longitudinal and transverse polarizations. The suppression and broadening are direct signatures of accelerated quarkonium dissociation. A bigger $\mu $ corresponds to a denser medium, which can screen the color charge between the heavy quark pair more effectively and provide additional channels for inelastic scattering, thereby destabilizing the bound state.

Fig.~\ref{fig:2} conclusively demonstrates that an increasing chemical potential promotes the dissociation of both $J/\Psi $ and $\Upsilon (1S)$, complementing the thermal effects shown in Fig.~\ref{fig:1}. The findings reinforce the picture that a hotter and denser QGP is more effective in dissociating heavy quarkonium states. This analysis, combined with the anisotropic and rotating framework, provides essential insights for interpreting quarkonium dissociating data occurring under complex conditions. This is also consistent with previous studies that have manifested that finite temperature and density effects enhance the melting of heavy quarkonium states~\cite{Braga:2017bml,Bao:2024glw,Mamani:2022qnf,Braga:2017oqw}.
\begin{figure}[H]
	\centering
	\includegraphics[width=0.62\textwidth]{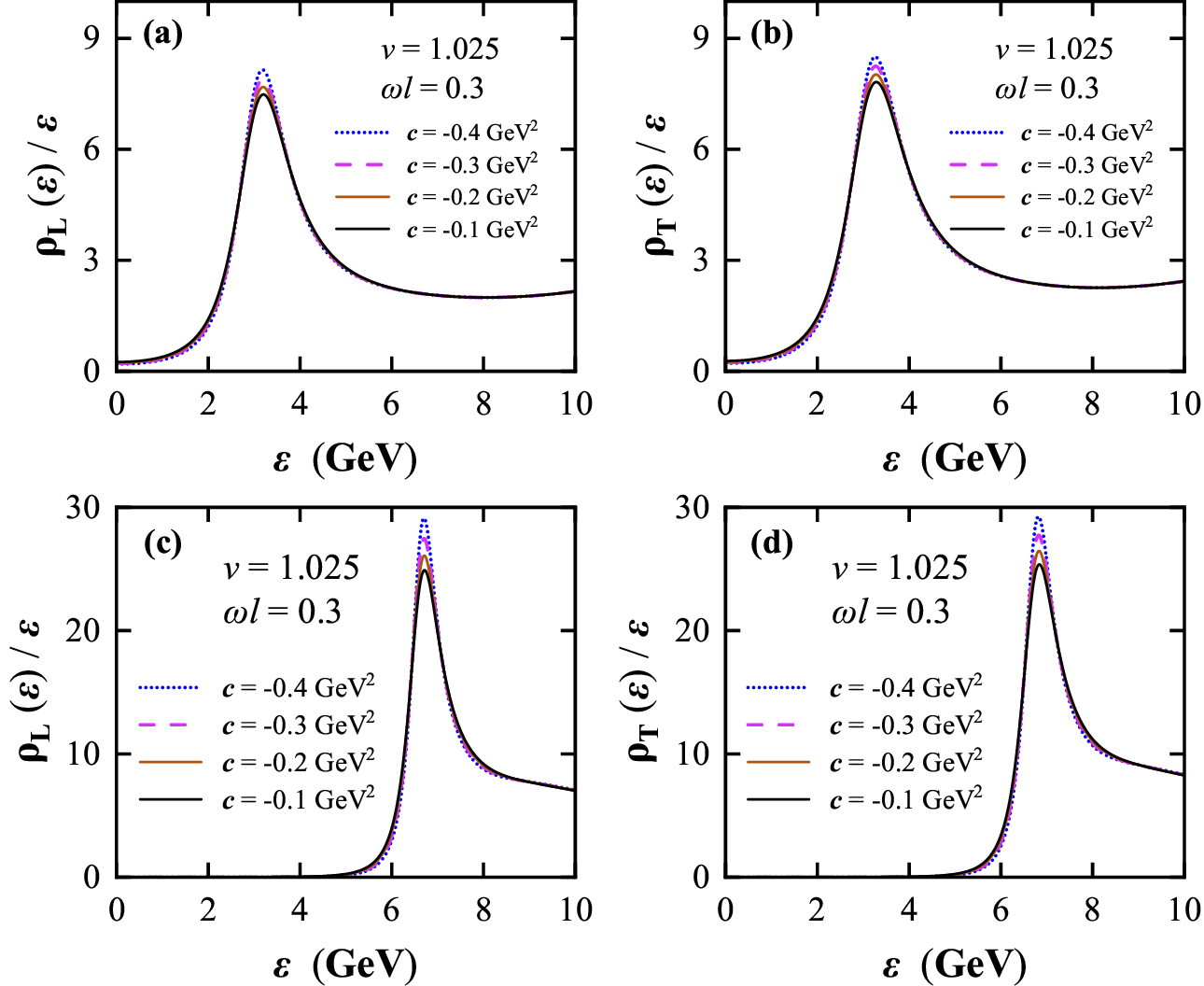}
	\caption{Spectral functions of $J/\Psi $ and $\Upsilon (1S)$ with varying warp factor coefficient, at fixed temperature, $T$ = 0.4 GeV, chemical potential $\mu  = 0.1 ~\textrm{GeV}$. Panels (a) and (b) correspond to $J/\Psi $,  (c) and (d) correspond to $\Upsilon (1S)$. For each meson, left/right panels show longitudinal/transverse polarizations relative to the anisotropic direction respectively.}
	\label{fig:3}
\end{figure}
Fig.~\ref{fig:3} investigates the effect of the geometric warp factor coefficient $c$ which signifies the deviation from conformality on heavy quarkonium spectral functions. As can be seen from Fig.~\ref{fig:3} , increasing $c$ leads to clear peak suppression and width broadening for both $J/\Psi $ and $\Upsilon (1S)$, in both longitudinal and transverse channels. This indicates that an increase in the warp factor coefficient facilitates the dissociation of heavy quarkonia.

The warp factor controls a fundamental deformation of the bulk spacetime geometry in the holographic dual. Its effect demonstrates that the stability of heavy quarkonia is sensitive not only to standard thermodynamic parameters ($T,\mu $) but also to the underlying geometric structure of the strongly coupled medium. Enhanced deformation likely increases the effective string tension or energy required to maintain the holographic string dual to the quarkonium state, which promotes its breakup. This result underscores that in a dynamically evolving QGP, changes in the medium's geometry can directly contribute to quarkonium melting, showing a purely geometric mechanism to the dissociation processes.

\subsection{THE INFLUENCE OF ROTATION AND ANISOTROPY ON THE SPECTRAL FUNCTIONS }\label{sec:B}
\begin{figure}[H]
	\centering
	\includegraphics[width=0.62\textwidth]{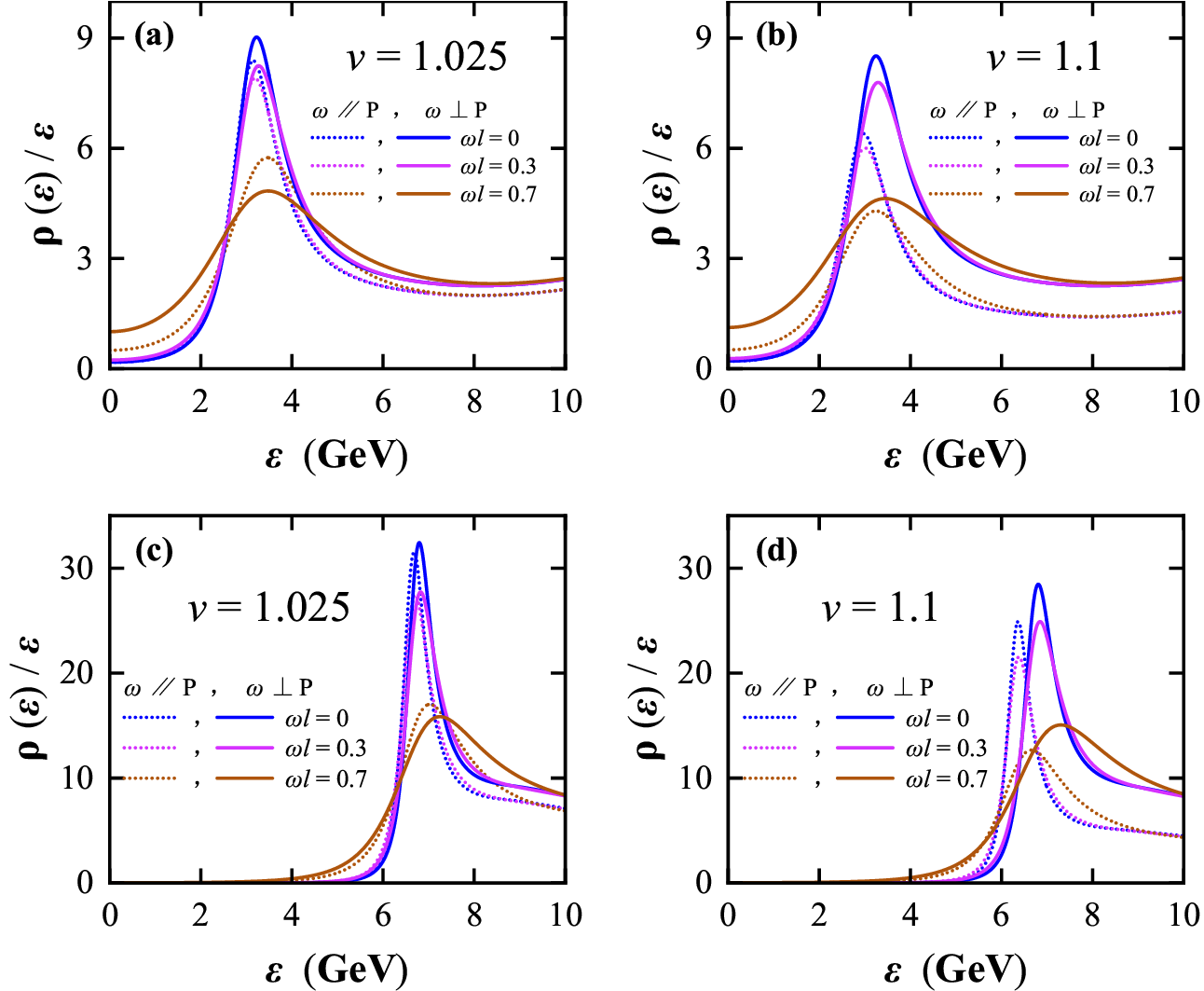}
	\caption{Spectral functions of $J/\Psi $ and $\Upsilon (1S)$ at $\mu  = 0.1\,{\rm{GeV}},c =  - 0.3\;{\rm{Ge}}{{\rm{V}}^2}$ and $T$ = 0.4 GeV under varying angular velocities, comparing two anisotropy strengths: (a, c) correspond to smaller anisotropy ($\nu  = 1.025$); (b, d) correspond to larger anisotropy ($\nu  = 1.1$). Panels (a, b) show $J/\Psi $, and panels (c, d) show $\Upsilon (1S)$. Left/right sides within each meson correspond to longitudinal/transverse polarizations relative to the anisotropic direction.}
	\label{fig:4}
\end{figure}
Fig.~\ref{fig:4} presents a central finding of this work: the competitive interplay between global rotation and spatial anisotropy in dissociating heavy quarkonia. Two key observations emerge: (1). both increasing angular velocity ($\omega l$) and anisotropy parameter ($\nu $) suppress spectral peaks and broaden widths for $J/\Psi $ and $\Upsilon (1S)$, confirming each factor independently enhances quarkonium melting in the QGP.  (2). The relative impact of rotation and anisotropy depends critically on their strengths: At small anisotropy ($\nu  = 1.025$), rotation's influence grows with angular velocity. At small angular velocity, anisotropy slightly dominates (longitudinal peaks lower). At big angular velocity, rotation becomes dominant (longitudinal peaks higher than transverse ones). At large anisotropy ($\nu  = 1.1$), anisotropy overwhelmingly dominates the dissociation pattern regardless of whether the angular velocity is large or small values, with longitudinal peaks remaining significantly more suppressed.

This competition mirrors the complex early-time dynamics in non-central heavy-ion collisions, where an initially anisotropic QGP also carries substantial angular momentum. The results indicate that the dominant mechanism quenching quarkonium depends on the collision's impact parameter and evolution stage. A strong anisotropic initial state (large $\nu $) may imprint a stronger directional suppression signature, when in scenarios with weaker initial anisotropy but significant vortex development, rotational effects could govern the dissociation dynamics, particularly for transversely polarized states. This nuanced understanding is crucial for interpreting polarization-dependent quarkonium suppression data, as it disentangles the concurrent effects of spatial deformation and collective rotation in the created fireball.

\begin{figure}[H]
	\centering
	\includegraphics[width=0.62\textwidth]{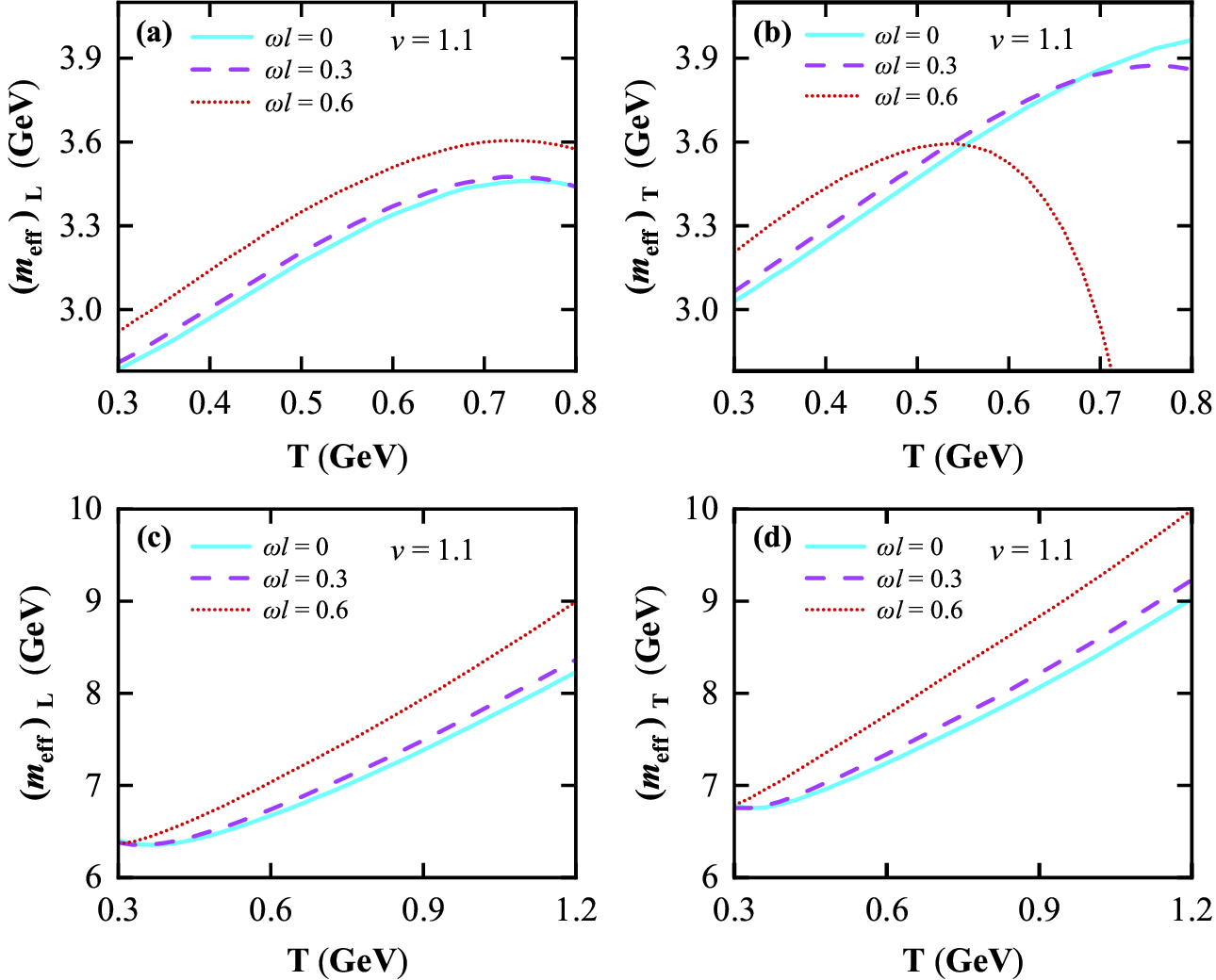}
	\caption{The effective mass of $J/\Psi $ and $\Upsilon (1S)$ as a function of temperature with different angular velocities at $\mu  = 0.1\, {\rm{GeV}}, c =  - 0.3\;{\rm{Ge}}{{\rm{V}}^2}$ and $\nu  = 1.1$. Panels (a) and (b) are the effective mass of  $J/\Psi $, and panels (c) and (d) are the effective mass of $\Upsilon (1S)$. (a) and (c) are parallel to the anisotropic direction, and (b) and (d) are perpendicular to the anisotropic direction.}
	\label{fig:5}
\end{figure}
Fig.~\ref{fig:5} shows the effective mass of $J/\Psi $ and $\Upsilon (1S)$ as a function of temperature with different angular velocities at $\mu  = 0.1\,{\rm{GeV}},c =  - 0.3\; {\rm{Ge}}{{\rm{V}}^2}$ and $\nu  = 1.1$, Figs.~\ref{fig:5}(a) and (b) present the effective masses of $J/\Psi $, and Figs.~\ref{fig:5}(c) and (d) present those of $\Upsilon (1S)$.

As can be seen from Fig.~\ref{fig:5}, for $J/\Psi $, increasing rotation angular velocity increases the effective mass in both longitudinal and transverse channels but it induces a non-monotonic behavior only in the transverse direction. The non-monotonic behavior in the longitudinal direction is mainly caused by anisotropy. This reflects how anisotropy softens the medium along its derection, whereas rotation generates centrifugal effects in the transverse direction. For $\Upsilon (1S)$, rotation monotonically increases the effective mass in both directions, but it does not induce non-monotonic behavior, which demonstrates compared to $J/\Psi $, $\Upsilon (1S)$ has stronger binding interactions. Another noteworthy and interesting point is that, for both $J/\Psi $ and $\Upsilon (1S)$, the longitudinal effective mass consistently remains smaller than the transverse effective mass. These results reveal that the in-medium mass shift is not merely thermal but geometrically and dynamically modulated by the plasma's global rotation and anisotropy, offering a finer interpretation of polarization-dependent quarkonium signatures in non-central heavy-ion collisions.
\begin{figure}[H]
	\centering
	\includegraphics[width=0.62\textwidth]{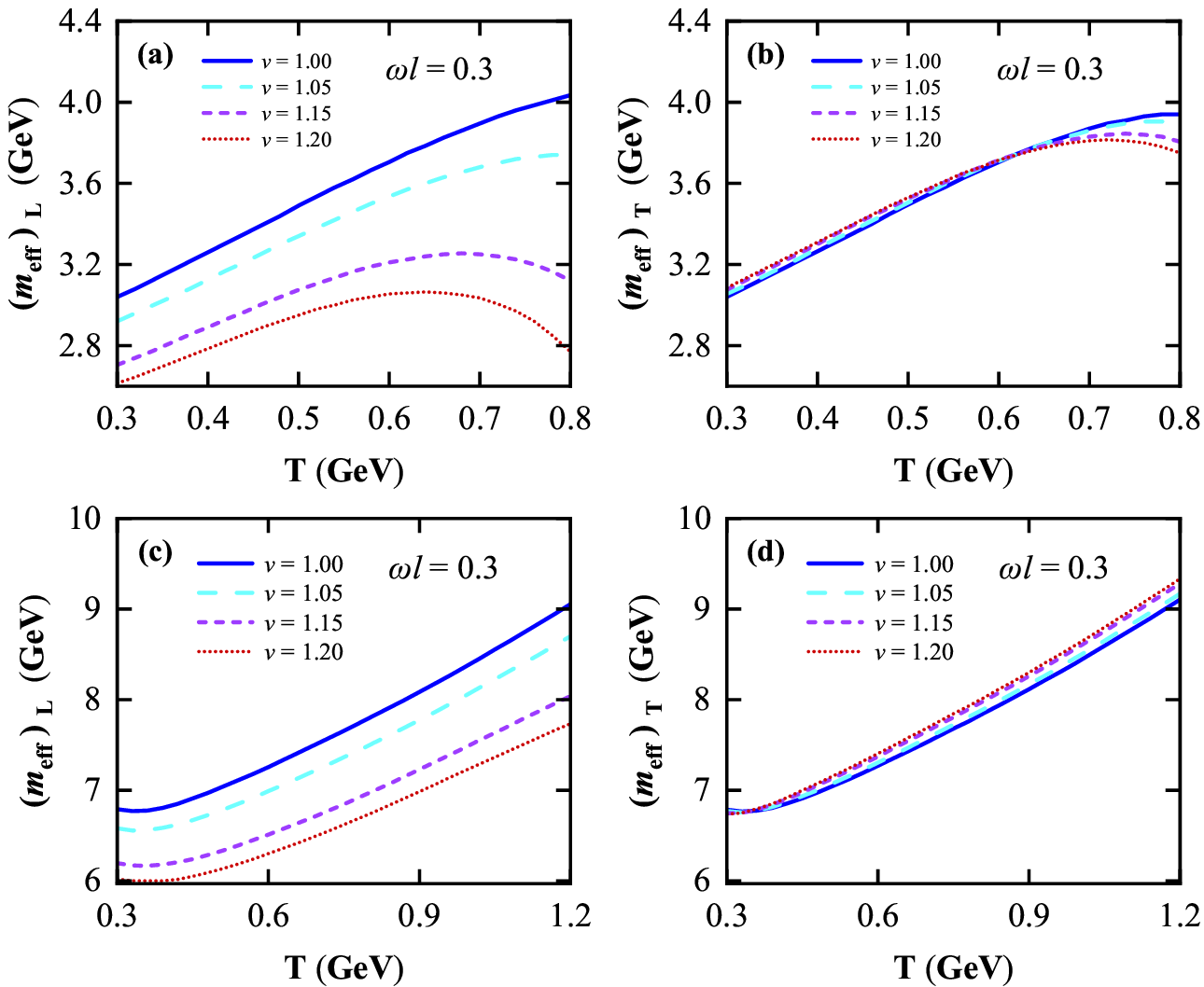}
	\caption{The effective mass of $J/\Psi $ and $\Upsilon (1S)$ as a function of temperature with different anisotropy  at $\mu  = 0.1\,{\rm{GeV}},\;c =  - \,0.3\;{\rm{Ge}}{{\rm{V}}^2}$ and $\omega l = 0.3$. Panels (a) and (b) are for the effective mass of $J/\Psi $, and panels (c) and (d) are for the effective mass of $\Upsilon (1S)$. (a) and (c) are parallel to the anisotropic direction, and (b) and (d) are perpendicular to the anisotropic direction.}
	\label{fig:6}
\end{figure}
Fig.~\ref{fig:6} details the anisotropic modulation of effective masses for $J/\Psi $ and $\Upsilon (1S)$ at a fixed rotational angular velocity ($\omega l = 0.3$). For $J/\Psi $, increasing the anisotropy parameter $\nu$ significantly suppresses the longitudinal effective mass and induces a non-monotonic temperature dependence. In the transverse channel, anisotropy slightly enhances the effective mass at low temperatures but reduces it at higher temperature. This directional effection highlights how anisotropy distinctively softens the medium along its anisotropic direction. which also explains why the longitudinal effective mass is smaller than the transverse effective mass in Fig.~\ref{fig:5} . For the more tightly bound $\Upsilon (1S)$, anisotropy strongly reduces the longitudinal mass but only marginally increases the transverse mass, underscoring its resilience to transverse deformation.

The underlying physics reveals a clear geometric influence: anisotropy stretches the medium along one direction, weakening the color-confining potential for mesons polarized parallel to it, thereby lowering their effective mass. Perpendicularly polarized states experience a compressed, stiffer environment, which can initially support a slightly higher mass until thermal effects dominate at high temperature. Crucially, when combined with the rotational effects shown in Fig.~\ref{fig:5}, Fig.~\ref{fig:6} demonstrates that anisotropy and rotation are independent but interactive mechanisms that directionally reshape the quarkonium in-medium spectrum. This provides a holographic explanation for polarization-dependent suppression patterns observed in non-central heavy-ion collisions, where the early-stage spatial deformation of the QGP imprints a specific directional signature onto heavy-flavor probes, complementing the later-stage rotational influence.

\section{SUMMARY AND CONCLUSIONS }\label{sec:5}
In this holographic study, we have investigated how global rotation and spatial anisotropy concurrently modify the in medium spectral functions and effective masses of heavy quarkonia ($J/\Psi $ and $\Upsilon (1S)$) in a strongly coupled quark gluon plasma. By employing a gauge/ gravity model that incorporates finite temperature, chemical potential, warp factor, and both rotational and anisotropic deformations, we obtain several key results that elucidate the interplay between collective dynamics and quarkonium dissociation in non central heavy ion collisions.

Increasing temperature (T), baryon chemical potential ($\mu$), anisotropy strength ($\nu $), or angular velocity ($\omega l$) consistently suppresses spectral peaks and broadens their widths for both charmonium and bottomonium. This demonstrates that quarkonium melting is accelerated not only by thermal agitation and density effects but also by geometric distortion and centrifugal forces induced by rotation. The underlying mechanism is that each factor independently weaken the color confining potential binding the heavy quark pair: higher T and $\mu$ enhance color screening and inelastic scatterings, while anisotropy stretches the medium and rotation imposes a velocity gradient that destabilizes bound states.
\begin{figure}[H]
	\centering
	\includegraphics[width=0.61\textwidth]{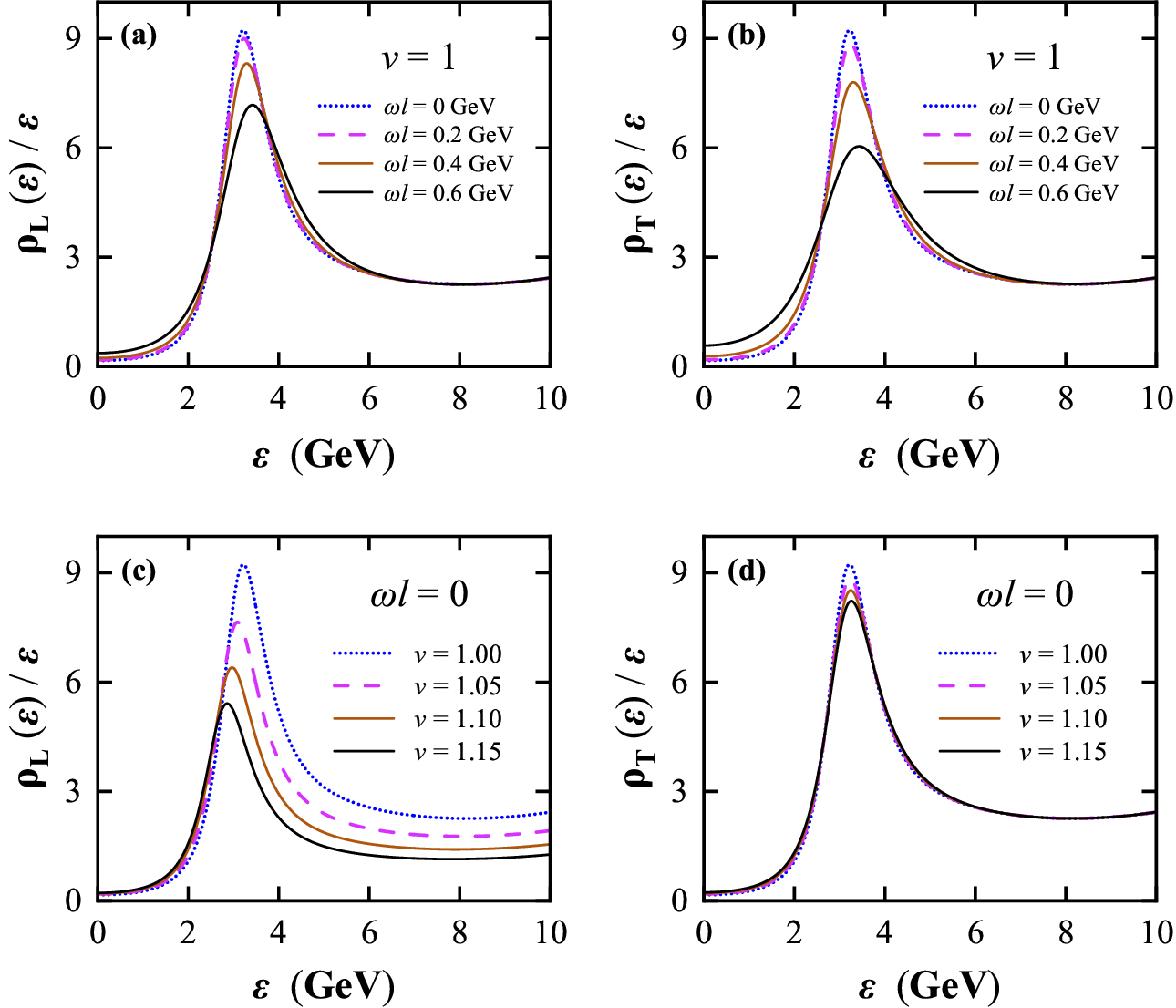}
	\caption{Spectral functions of $J/\Psi $ with varying angular velocity and anisotropies, at $T = 0.4~\textrm{GeV}$ $\mu  = 0.1~\textrm{GeV}$ and $c =  - 0.3~\textrm{GeV}{^2}$. The left/right panels show longitudinal/transverse polarizations relative to the anisotropic direction respectively.}
	\label{fig:7}
\end{figure}
To analyze the effects of both anisotropy and rotation on the dissociation of heavy vector mesons, there are differences along the dominant directions, we make a discussion of the influence of rotation on dissociation under isotropic conditions, as well as the influence of anisotropy on dissociation in the absence of rotation. The detailed results are shown in Fig.~\ref{fig:7}. From Figs.~\ref{fig:7} (a) and (b), it can be observed that at the same angular velocity, the peak of the spectral function in the transverse channel is significantly lower than that in the longitudinal channel, with a broader broadening, indicating that rotation exerts a stronger dissociation effect along the transverse direction. From Figs.~\ref{fig:7} (c) and (d), it can be seen that under the same anisotropic condition without rotation, the spectral function peak in the longitudinal channel is lower and exhibits broader broadening, demonstrating that anisotropy acts in the opposite manner to rotation$-\/$anisotropy has a stronger effect along the longitudinal direction. Anisotropy preferentially dissociates mesons polarized parallel to the anisotropy axis (longitudinal channel), as the medium's expansion along this direction weakens the binding force. In contrast, rotation more strongly affects mesons polarized perpendicular to the rotation axis (transverse channel) due to centrifugal stretching. This directional selectivity reflects the broken spatial symmetry of the QGP and provides a clear theoretical basis for interpreting polarization dependent suppression patterns in experiment.

The combined effect of rotation and anisotropy is not additive but competitive, depending on their relative strengths. For small anisotropy, rotation becomes the dominant dissociation mechanism at high angular velocities, whereas anisotropy governs at low rotation. For large anisotropy, anisotropy dictates the dissociation regardless of rotation strength. This competition mirrors the space time evolution of the fireball: initial anisotropy may imprint a directional suppression signature early on, while developed vorticity later governs dissociation, particularly for transverse polarizations. The result implies that quarkonium data at different collision centralities and energies can disentangle the relative contributions of initial anisotropy and rotation.

The effective masses of $J/\Psi $ displays the polarization dependent non monotonic behavior with temperature. For $J/\Psi $, rotation induces non monotonic variation in the transverse effective mass, while large anisotropy causes similar behavior in the longitudinal effective mass. This indicates that the in-medium mass shift is not purely thermal but is modulated by the dynamical and geometric distortions of the plasma.

In summary, this work provides a coherent non perturbative account of how rotation and anisotropy$-\/$two characteristics of the QGP in off-central collisions reshape the heavy quarkonium spectrum. The model offers a quantitative foundation for disentangling these effects in experimental data, with direct relevance to ongoing and future studies at RHIC and the LHC. Future extensions could incorporate time dependent anisotropy, more realistic collision profiles, and electromagnetic fields to further advance our understanding of heavy flavor dynamics in extreme QCD environments.

\begin{acknowledgements}
	This work was supported by the National Natural Science Foundation of China (Grants No. 12575144, and No. 11875178).
\end{acknowledgements}
\bibliography{Ref}	
\end{document}